\documentclass[article,twocolumn,amsmath,amssymb,noshowpacs,prb]{revtex4}
\usepackage{graphicx}
\usepackage{bm}
\usepackage{amsmath}
\usepackage{amsxtra}
\def\beq#1\eeq{\begin{equation}#1\end{equation}}
\def\beqry#1\eeqry{\begin{eqnarray}#1\end{eqnarray}}

\newcommand{\sml}[1]{ \scriptscriptstyle #1 }

\newcommand{\bsigma}{\mbox{\boldmath$\sigma$\unboldmath}}
\newcommand{\blambda}{\mbox{\boldmath$\lambda$\unboldmath} }

\newcommand{\calw}{\mathcal{W}}

\newcommand{\br}{{\bf r}}
\newfont{\ccc}{cmbxsl10 at 10pt}

\begin{document}

\title{
Magnetic Exchange Couplings 
from Noncollinear Spin Density Functional Perturbation Theory}
\author{Juan E. Peralta}
\author{Veronica Barone}
\affiliation{Department of Physics, Central Michigan University, Mt. Pleasant, MI 48859 }
\date{\today}

\begin{abstract}
We propose a method for the evaluation of   magnetic exchange couplings based on  
noncollinear spin-density functional calculations. 
The method employs the second derivative of the total Kohn-Sham energy of a single reference state, in contrast  to
approximations based on Kohn-Sham total energy differences.
The advantage of our approach is twofold: 
It provides a physically motivated picture of the transition from a low-spin to a high-spin state, 
and it utilizes a  perturbation scheme for the evaluation of magnetic exchange couplings. 
The latter  simplifies the way these parameters are predicted using first-principles:
It avoids the non-trivial search for different spin-states that  needs to be carried out in energy difference methods
and it opens the possibility of ``black-boxifying'' the extraction
of exchange couplings  from density functional theory calculations. 
We  present  proof of concept calculations of magnetic exchange couplings in 
the H--He--H model system and in an oxovanadium bimetallic complex where the results 
can be intuitively rationalized.
\end{abstract}


\maketitle

\section{Introduction}

Empirical models based on the Heisenberg spin Hamiltonian are routinely  utilized
to describe the behavior of a variety of magnetic systems.  
In most cases, these simple models are found to fit the
experimental data very well, provided that the parameters in the model Hamiltonian are chosen properly.
The set of parameters can include both, external parameters (temperature, applied magnetic field, etc.), and internal
parameters (magnetic exchange couplings, magnetic anisotropy, etc.).
Internal parameters for a particular system can be obtained either by fitting experimental data or 
from first-principles electronic structure calculations by 
 mapping total electronic energies to the energies of the Heisenberg spin Hamiltonian.\cite{Noodleman_JCP_74_5737,Ruiz_JACS_119_1297,Kortus_Polyhedron_22_1871}
In particular, magnetic exchange
couplings, $J$, can be obtained considering the isotropic Heisenberg Hamiltonian
\begin{equation}
\label{eq:Heisenberg}
\hat{H} = - 2 \sum_{<i,j>} J_{ij} \, \hat{\mathbf{S}}_i \cdot \hat{\mathbf{S}}_j  \,,
\end{equation}
where $\hat{\mathbf{S}}_i$ and  $\hat{\mathbf{S}}_j$ are the (localized) spin operators associated to each magnetic center.

Perhaps the one of the most interesting manifestation of magnetism at the molecular scale can be found in 
complexes containing transition metal atoms. 
Many applications have been suggested  exploiting these molecular-size magnets, such as quantum
computation units and high-density data storage.\cite{Christou_MRS} 
Due to the relatively large size of most complexes of interest, density functional theory (DFT)\cite{kohn64,vonBarth_Hedin}
offers the most efficient alternative for modeling the electronic structure of these systems 
from first-principles.\cite{Kortus_PRL_86_3400,Kortus_Polyhedron_22_1871,Ribas_JCP_123_044303}

Several approaches had been proposed to  extract $J$ couplings from DFT energies. According to the spin-projected (SP)
approach,\cite{Noodleman_JCP_74_5737,Noodleman_AIC_38_423,Noodleman_CP_109_131} the energies of a two-center complex $A$ and $B$ can be related to the $J$ coupling as
\begin{equation}
\label{eq:SP}
E_{\sml LS} - E_{\sml HS} = 4 S_A S_B  J_{\sml AB}  \,,
\end{equation}
while in the non-projected (NP) approach\cite{Ruiz_JCC_20_1391}, 
the energies of a two-center complex $S_A$ and $S_B$ can be related to the $J$ coupling as
\begin{equation}
\label{eq:NP}
E_{\sml LS} - E_{\sml HS} = (4 S_A S_B + 2 S_B) J_{\sml AB} \,,
\end{equation}
where $S_B \le S_A$.
In Eqs.~(\ref{eq:SP}) and (\ref{eq:NP}), $E_{\sml HS}$ is the energy of the high-spin state and $E_{\sml LS}$ is the energy of the
low-spin (broken-symmetry)
state. Eqs.~(\ref{eq:SP}) and (\ref{eq:NP}) can be straightforwardly generalized to a set of equations for complexes
with multiple magnetic centers.\cite{Noodleman_JACS_107_3418,Ruiz_JCC_24_982} While the SP and NP methods
are fairly popular, other methods have been proposed in the literature such as Nishino's approach\cite{Nishino_JPCA_101_705}, the constrained-DFT
approach of Rudra et al.,\cite{Rudra_JCP_124_024103,Rudra_IC_46_10539}, the Slater's transition state method of Dai and  Whangbo.\cite{Dai_JCP_114}
and the local spin method of Clark and Davidson.\cite{Clark_JCP_115_7382,Davidson_JPCA_106_7456}
All these approaches rely on the evaluation of the energy difference between two (for the simplest case of a bimetallic complex)  or more states.
The evaluation of this energy difference  is commonly done by carrying out several self-consistent field  calculations, one for each different magnetic configuration. 
However, in many cases, converging to the right target state could be 
cumbersome, specially for  systems containing multiple centers with many magnetic configurations. Therefore, developing an approach that can be used in a  ``black-box''  manner 
is crucial to systematically  explore a large set of complexes or complexes containing many magnetic centers. 

In this work, we present an approach for the evaluation of magnetic exchange couplings based on noncollinear spin density 
functional calculations that allows, in analogy to response properties, to express the magnetic 
exchange couplings as a derivative of the total electronic energy of one single state with respect to an external parameter, 
 opening the possibility of ``black-boxifying'' the extraction 
of magnetic exchange couplings from density functional theory calculations.

\section{Theory and Implementation}

\subsection{Exchange Couplings as Energy Derivatives}

Let us consider the effective interaction energy between two  magnetic centers $A$ and $B$  given by
the isotropic classical Heisenberg model (obtained by taking the expectation value of Eq.~(\ref{eq:Heisenberg})), 
\begin{eqnarray}
\label{eq:Heis-cosine}
E_{\sml AB} & = & - 2 J_{\sml AB} \, \mathbf{S}_A \cdot \mathbf{S}_B \nonumber \\
            & = & - 2 J_{\sml AB} \, S_A S_B \cos \theta  \,,
\end{eqnarray}
where $\mathbf{S}_A$  and $\mathbf{S}_B$  are the 
(perfectly localized) magnetic moment vectors, $J_{\sml AB}$ is the exchange coupling constant, and $\theta$ is
the angle between $\mathbf{S}_A$ and $\mathbf{S}_B$. From Eq.~(\ref{eq:Heis-cosine}), one can trivially obtain 
$J_{\sml AB}$ from the second derivative of $E_{\sml AB}$ with respect to $\theta$ at the equilibrium points,
\begin{eqnarray}
\label{eq:Heis-deriv}
J_{\sml AB} & = &   \frac{1}{2 S_A S_B} \left( \frac{d^2 E_{\sml AB}}{d\theta^2} \right)_{\theta=0}  \, \nonumber \\
       & = & - \frac{1}{2 S_A S_B} \left( \frac{d^2 E_{\sml AB}}{d\theta^2} \right)_{\theta=180^\circ}\,.
\end{eqnarray}
These simple relations provide a direct path to the  evaluation of magnetic exchange couplings $J_{\sml AB}$ 
from density functional calculations if $E_{\sml AB}$ in Eq.~(\ref{eq:Heis-deriv}) is replaced by the total Kohn-Sham  (KS)\cite{kohn65} energy of the system, $E^{\sml KS}$. 
Therefore, assuming that the electronic system  depends on $\theta$ as an ideal Heisenberg model (the validity of this 
assumption will be discussed in the next Section), one can express the exchange coupling constant $J_{\sml AB}$ 
in terms of an energy derivative as
\begin{eqnarray}
\label{eq:Heis-deriv1}
J_{\sml AB} & = &   \frac{1}{2 S_A S_B} \left( \frac{d^2 E^{\sml KS}}{d\theta^2} \right)_{\theta=0}  \, 
\end{eqnarray}
or
\begin{eqnarray}
\label{eq:Heis-deriv2}
 J_{\sml AB}     & = & - \frac{1}{2 S_A S_B} \left( \frac{d^2 E^{\sml KS}}{d\theta^2} \right)_{\theta=180^\circ}\,,
\end{eqnarray}
where the angle $\theta$ in the DFT framework is defined as the angle between the 
local magnetization vectors $\mathbf{S}_A$  and  $\mathbf{S}_B$. 
Another related method based on the Green's function formalism for  
crystals has been proposed by Liechtenstein et al.\cite{Liechtenstein} 

\subsection{Constraint Noncollinear Spin-DFT Calculations}

To evaluate the dependence of $E^{\sml KS}$ on $\theta$, we first introduce
two-component spinors as Kohn-Sham  orbitals,
\begin{eqnarray}
\Psi_i(\br)  = \left(
\begin{array}{c}
\psi_i^{\uparrow} (\br) \\
\psi_i^{\downarrow} (\br)
\end{array}
\right)
\label{Eq:Psi}
\,,
\end{eqnarray}
where $\psi_i^\uparrow (\br)$ and $\psi_i^\downarrow (\br)$ are spatial orbitals 
expanded in a linear combination of atomic orbitals,
\begin{equation}
\psi_i^\omega({\bf r}) = \sum_\mu c_{\mu i }^{\omega} \phi_\mu
({\bf r})~~~~~~~~~~~ (\omega = \uparrow, \downarrow)
\label{Eq:Spinors}
\,.
\end{equation}
The two-component spinors introduce the freedom in the spin-dependence of the KS system that allows for
 local rotations of the spin density characterized by $\theta\ne0$ and $\theta\ne180^\circ$, i.e. noncollinear 
spin densities.\cite{Kubler_1988,Nordstrom_1996,Oda_1998,Kurz_2004,Yamanaka_2000}
The local magnetization vectors  $\mathbf{S}_A$  and  $\mathbf{S}_B$ can be written as
\begin{eqnarray}
\mathbf{S}_{A,B} = \int\,d^3r\, \calw_{A,B}(\br) \mathbf{s}(\br)\,,
\end{eqnarray}
where 
\begin{eqnarray}
\label{eq:spin}
\mathbf{s}(\br) = \sum_{i \in occ} \Psi_i^\dagger ({\br}) \bsigma \Psi_i({\bf r})
\end{eqnarray}
is the spin-density vector and $\calw_{A,B}(\br)$ is a scalar weight function that determines each local magnetic site.
It is important to recall that the magnetic centers $A$ and $B$ represent a group of one or more atoms.
In Eq.~(\ref{eq:spin}), $\bsigma =(\sigma_x, \sigma_y, \sigma_z)$ is the $2\times 2$ Pauli matrices vector.

\begin{figure}[h]
   \includegraphics[width=5.3cm]{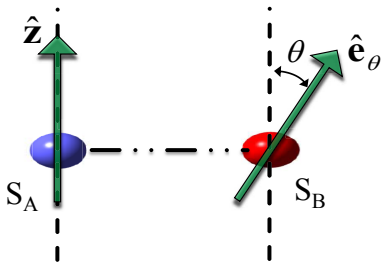}
\caption{Schematic representation of the constraint vectors employed for the local rotations of the spin density. }
  \label{Fig:scheme}
\end{figure}

Having defined the local magnetic moments, the second step is to find the dependence of the total electronic energy upon
local rotations of the spin density. This is done by constraining the {\em direction} of the local
magnetizations  $\mathbf{S}_A$  and  $\mathbf{S}_B$  by means of the Lagrange multipliers technique. 
To this end, we construct the  Lagrangian functional $\Lambda$
\begin{eqnarray}
\label{eq:Lagrangian}
\Lambda[\{\Psi({\br})\},\blambda_A,\blambda_B] =  E^{\sml KS}[\{\Psi({\br})\}] - \nonumber \\
\blambda_A \cdot (\mathbf{S}_A \times \hat{\bf z}) 
- \blambda_B \cdot (\mathbf{S}_B \times \hat{\bf e}_\theta)
\,,
\end{eqnarray}
where $\hat{\bf e}_\theta = \sin \theta \, \hat{\bf x} + \cos \theta \, \hat{\bf z}$ is a unity vector to which $\mathbf{S}_B$ is constraint to be parallel to, 
and for simplicity $\mathbf{S}_A$ has been
chosen to be constraint to the $z$ direction, as schematized in Fig.~\ref{Fig:scheme}. 
Here $0 \le \theta \le 180^\circ$ is considered as an external parameter for which $(dE/d\theta)_{\theta=0}=(dE/d\theta)_{\theta=180^\circ}=0$.
Eq.~(\ref{eq:Lagrangian}) can be readily generalized for the case of many magnetic 
centers and arbitrary unity vector directions. For the case of two magnetic centers 
(and for the purpose of this work), Eq.~(\ref{eq:Lagrangian}) does not imply any loss of generality.
In Eq.~(\ref{eq:Lagrangian}), $E^{\sml KS}[\{\Psi({\br})\}]$ represents the KS energy
of the system which is, in practice, a functional of the set of occupied KS orbitals $\{\Psi({\br})\}$. 
Stationary solutions of $\Lambda$ for a given (fixed) $\theta$ imply
\begin{eqnarray}
\label{eq:stat1}
\frac{d \Lambda}{d \blambda_A } = \mathbf{S}_A \times \hat{\bf z} = \bf{0}\,,
\end{eqnarray}
\begin{eqnarray}
\label{eq:stat2}
\frac{d \Lambda}{d \blambda_B } = \mathbf{S}_B \times \hat{\bf e}_\theta = \bf{0}\,,
\end{eqnarray}
and
\begin{eqnarray}
\label{eq:stat3}
& & \frac{\delta \Lambda}{\delta \Psi_i^\dagger({\br}) } =\frac{\delta E^{\sml KS}}{\delta \Psi_i^\dagger({\br})} - 
\Big[\calw_{A}(\br)\blambda_A\cdot(\bsigma\times\hat{\bf z})+ \nonumber  \\
& & \;\;\;\;\;\calw_{B}(\br)\blambda_B
\cdot(\bsigma\times\hat{\bf e}_\theta)\Big]\Psi_i({\bf r}) = 0 ~~~~~~~~~~~ (i \in \mbox{occ}).
\end{eqnarray}
While Eqs.~(\ref{eq:stat1}) and (\ref{eq:stat2}) restore the constraint conditions, Eq.~(\ref{eq:stat3}) combined with the orthonormality 
condition for the spinors yields a modified set of  KS equations (in terms of two-component spinors) that include 
the two additional terms inside the square brackets on the left-hand side of Eq.~(\ref{eq:stat3}),
\begin{eqnarray}
\label{eq:KS}
\Big[ T + V_N + J + V_{xc} - \calw_{A}(\br) \blambda_A \cdot (\bsigma \times \hat{\bf z} ) - \nonumber \\
 \calw_{B}(\br) \blambda_B \cdot( \bsigma \times \hat{\bf e}_\theta ) 
 \Big]   \Psi_i ({\bf r})  = \epsilon_i \Psi_i ({\bf r})   \,,
\end{eqnarray}
where $T=-1/2\,\nabla^2$ is the kinetic energy, $V_N$ is the electron-nuclei potential,
$J$ is the Coulomb (or Hartree) potential, and $V_{xc}$ is the exchange-correlation (XC) potential.
The sum of the first four terms inside the square brackets is the standard KS Hamiltonian, while the two  additional terms
can be interpreted as a potential originated in a torque exerted on the local magnetic moments $\mathbf{S}_A$ and $\mathbf{S}_B$.
It should be noted that other approaches had been proposed in the literature 
to constraint the direction of the local mangetization in spin DFT calculations.\cite{Dederichs_1984,Sticht_1989,sandratskii_1988}

It is important to note that since the constraint conditions are linear in the spin density 
vectors, the additional terms in Eq.~(\ref{eq:KS})  depend implicitly on the orbitals only through $\blambda_A $ and $\blambda_B$, which simplifies 
the implementation.

Neglecting spin-orbit interaction,  $T$, $V_N$, and $J$ are diagonal in the $2 \times 2$ spin space, and thus
the only term in Eq.~(\ref{eq:KS}) that couples $\psi_i^{\uparrow}$ and  $\psi_i^{\downarrow}$  is $V_{xc}$. 
In a previous work, we have generalized  $V_{xc}$ for noncollinear magnetizations,\cite{Peralta_PRB_75_125119} 
assuming that  the XC energy depends on the local variables  in the same manner as in
the standard collinear (spin-unrestricted) case, and imposing the condition for the XC energy to be  invariant under rigid rotations of
the spin density. In that work, we have derived  $V_{xc}$  for general energy functionals containing a variety of ingredients 
 beyond the local-spin density approximation (LSDA) and the generalized-gradient approximation (GGA), 
such as meta-GGAs and  hybrid density functionals.
This same generalization is adopted throughout this work.
Other implementations  based on plane-waves\cite{Kubler_1988,Nordstrom_1996,Oda_1998,Kurz_2004} and Gaussian-type orbitals 
can be found in the literature for a variety of applications.\cite{Yamanaka_2000,Mayer_2001,Peralta_2004,Wang_2004,Malkin_2005,Armbruster_2008}

The constraint vectors can be chosen without loss of  generality to lay in the $x-z$ plane. 
Hence, as  spin-orbit interaction is not included in the Hamiltonian, the two-component spinors are purely real. 
As a consequence, in this scheme 
the orbital magnetization of the solutions is always  zero.

The value of $\Lambda$ at the stationary solutions for a fixed $\theta$ given by Eqs.~(\ref{eq:stat1})--(\ref{eq:stat3}) can be directly associated to
the energy of the KS system in the presence of the constraints, $E^{\sml KS}(\theta)$. 
The only arbitrariness in the formulation is the choice of the weight  function $\calw_{A,B}(\br)$ and the fragments $A$ and $B$.
Since in practice it is necessary to evaluate the matrix elements of $\calw_{A,B}(\br)$ in the  
atomic orbitals basis set $\phi_\xi ({\bf r})$, it is convenient to define $\mathbf{S}_A$ and $\mathbf{S}_B$ using population analysis,
\begin{eqnarray}
\label{eq:S-pop}
\mathbf{S}_{A,B} = \sum_{\mu,\nu} (\calw_{A,B})_{\mu\nu} \mathbf{P}_{\mu\nu} \,,
\end{eqnarray}
where $\mathbf{P}_{\mu\nu} $ is the spin-density matrix vector whose Cartesian components are
\begin{eqnarray}
\label{eq:P-matrix}
P^x_{\mu\nu} &    = & P^{\uparrow \downarrow}_{\mu\nu}  + P^{ \downarrow \uparrow}_{\mu\nu} \,, \\
P^y_{\mu\nu} &    = &  i ( P^{\uparrow \downarrow}_{\mu\nu} -  P^{\downarrow \uparrow}_{\mu\nu} ) \,,
\end{eqnarray}
and
\begin{eqnarray}
\label{eq:P-matrix2}
P^z_{\mu\nu} &   =  & P^{\uparrow \uparrow}_{\mu\nu}  + P^{\downarrow \downarrow}_{\mu\nu} \,,
\end{eqnarray}
written in terms of the $2 \times 2$ density matrix
\begin{eqnarray}
\label{eq:P-general}
P^{\omega\omega'}_{\mu\nu} = \sum_{i \in  \mbox{occ}} c_{\mu i }^\omega c_{\nu i }^{\omega'  \ast}  
\;\;\;\;\;\; (\omega, \omega' = \uparrow, \downarrow)    \,.
\end{eqnarray}
For this work, we employ L\"{o}wdin population analysis.\cite{Lowdin}  From the expression for the atomic spin-population given
by Eq.~(\ref{eq:S-pop}), $(\calw_{A,B})_{\mu\nu}$ can be obtained as
\begin{eqnarray}
(\calw_{A,B})_{\mu\nu}  = \frac{d S^\eta_{A,B}}{d P^\eta_{\mu\nu}} \;\;\;\;\;\;\;\; (\eta = x,y \mbox{~or~} z) \,.
\end{eqnarray}
Thus, using the L\"{o}wdin  partitioning, it is straightforward to show that the  matrix elements of $\calw_{A,B}(\br)$  can be calculated from the
atomic orbitals overlap matrix $(\mathbb{S})_{\lambda\sigma} = \int d^3r \phi_\lambda ({\bf r}) \phi_\sigma ({\bf r})$ as
\begin{eqnarray}
\label{eq:Lowdin}
 (\calw_{A,B})_{\mu\nu}  = \sum_{\lambda \in A,B} 
(\mathbb{S}^{1/2})_{\lambda\mu} (\mathbb{S}^{1/2})_{\nu\lambda}\,. 
\end{eqnarray}
It is worth commenting that atomic spin-densities are less sensitive to the choice of the population method than atomic densities.\cite{Ruiz_CCR_2005}
Since only the direction of the atomic spin-density is relevant in our method for the calculation of magnetic exchange couplings, 
we expect even less sensitivity with the coice of the population method.

The set of spinors that  simultaneously satisfy Eq.~(\ref{eq:KS}) and Eqs.~(\ref{eq:stat1}) and (\ref{eq:stat2})  needs to be 
determined self-consistently since $J$ and $V_{xc}$  depend on the spinors.  To obtain the KS Hamiltonian, we add the additional 
constraint terms of Eq.~(\ref{eq:KS}) to the standard KS Hamiltonian, initially
using a guess for the Lagrange multipliers $\blambda_A$ and $\blambda_B$. Then we determine the optimal $\blambda_A$ and $\blambda_B$ such that 
Eqs.~(\ref{eq:stat1}) and (\ref{eq:stat2}) are satisfied, using the density matrix obtained from diagonalizing the KS Hamiltonian of the constraint 
system to evaluate $\mathbf{S}_A$ and $\mathbf{S}_B$. 
This is carried out using the steepest descent method to find the minimum of $\Lambda$ as a function of $\blambda_A$ and $\blambda_B$. Once the optimal values of
  the Lagrange multipliers are determined, 
and provided that self-consistency is not achieved,
we proceed to the next self-consistent iteration using the density matrix from the previous iteration. The process stops once the criteria for 
changes in the density matrix and total energy are met.
Several consistency checks were performed to verify the robustness of our code.
We have implemented this scheme in the Gaussian Development Version program.\cite{gdv-f2}

Using this methodology, $E^{\sml KS}(\theta)$ was calculated for small values of $\theta$ around $\theta=0$ and $\theta=180^\circ$ and 
magnetic exchange couplings were obtained from the quadratic coefficient of a polynomial fit.
It is important  at this point to mention that there are cases where 
existing approximate density functional methods have difficulties
 in representing the LS state ($\theta=180^\circ$).\cite{Noodleman_JCP_74_5737,Ruiz_JCC_20_1391,Rudra_JCP_124_024103} 
The Kohn-Sham determinant in these cases correspond  to a ``broken-symmetry'' solution
that mixes two or more eigenfunctions of the $S^2$ operator. 
However, for the HS state ($\theta=0$),
it is customary accepted that approximate density functionals provide a reliable representation. 
Thus, even though for comparison purposes in the next Section we show our results using both the HS and LS states as reference, for the practical 
extraction of magnetic 
exchange couplings in the DFT framework, this method is expected to work more reliably using only the HS as the reference state.

%

\section{Proof of Concept Calculations}

We first tested our methodology in the H--He--H linear model system with a distance H--He of 1.625\AA,
considering  the outer H atoms as magnetic centers $A$ and $B$ ($S_A=S_B=1/2$).  
In Table~\ref{Table:HHeH} we show our  results for the magnetic exchange couplings calculated from $d^2E^{\sml KS}/d\theta^2$ 
using the $\theta=0$ ($\langle S_z \rangle = 1$) state and the $\theta=180^\circ$ ($\langle S_z \rangle = 0$) state, $J^{\sml HS}$ and $J^{\sml LS}$, respectively. 
All calculations were carried out with the 6-311G** Gaussian basis set.\cite{6311G} 
For comparison, in Table~\ref{Table:HHeH} we include results for the LSDA  (Dirac exchange plus the parametrization of Wosko, 
Wilk, and Nusair\cite{vosko80} for correlation), the BLYP  
realization of the GGA (Becke's 1988 functional\cite{becke88} for exchange and the correlation functional of Lee, Yang, and Parr\cite{lyp}), and for 
the B3LYP\cite{becke88,lyp,Becke_JCP_1993_B,B3LYP} hybrid functional. For all functionals, 
exchange couplings calculated from the energy derivatives, $J^{\sml HS}$ and $J^{\sml LS}$, are in very close agreement to the exchange coupling calculated 
from the energy difference, $J^{\sml \Delta E}$. The difference can be attributed to both, the intrinsic accuracy of 
the numerical differentiation method and to the fact that $J^{\sml HS}$ and $J^{\sml LS}$ are expected to be identical to $J^{\sml \Delta E}$ only in the case
where  DFT describes the electronic system as an ideal Heisenberg model. The small discrepancy between  $J^{\sml HS}$, $J^{\sml LS}$ and $J^{\sml \Delta E}$ can be understood in terms of the
 localized nature of the magnetization on the H atoms in this model system and provides a measure of 
how well the  electronic system mimics the behavior of an ideal Heisenberg model.

In Table~\ref{Table:HHeH}
we report $J^{\sml \Delta E}$ as calculated from the SP formula, Eq.~(\ref{eq:SP}), since it offers a direct comparison 
between $J^{\sml HS}$ and $J^{\sml LS}$, and $J^{\sml \Delta E}$.
It is worth mentioning that in the ideal case of 
a perfect Heisenberg system, $\Delta E = E(\theta = 180^\circ) - E(\theta = 0)$ 
is related to $J^{\sml HS}$ and $J^{\sml LS}$ according to
\begin{eqnarray}
\label{eq:relation}
\Delta E = 2   \left( \frac{d^2 E}{d\theta^2} \right)_{\theta=0} = - 2 \left( \frac{d^2 E}{d\theta^2} \right)_{\theta=180^\circ}  \,.
\end{eqnarray}
Therefore, $d^2 E/d\theta^2$ provides a measure of $\Delta E$ that can be evaluated without explicitly converging 
the self-consistent procedure to the LS state.

\begin{table}[h]
    \caption{
    Magnetic exchange couplings (in meV) calculated from energy derivatives and from energy differences for the 
H--He--H system. 
    \label{Table:HHeH}
    }
\begin{ruledtabular}
  \begin{tabular}{lrrr}
  & LSDA  & BLYP &  B3LYP  \\
\cline{1-4}
$J^{\sml HS} = \frac{1}{2 S_A S_B}   \left( \frac{d^2 E}{d\theta^2} \right)_{\theta=0}           $  & $-$95.8  &$-$74.0    &  $-$60.8  \\[6pt]
$J^{\sml LS} = - \frac{1}{2 S_A S_B} \left( \frac{d^2 E}{d\theta^2} \right)_{\theta=180^\circ} $    & $-$101.7  &$-$76.6   &  $-$61.2   \\[6pt]
$J^{\sml \Delta E}  =  \frac{E(\theta = 180^\circ) - E(\theta = 0)}{4 S_A S_B}$                           & $-$99.8  &$-$76.9   &  $-$63.5  \\[6pt]
    \end{tabular}
\end{ruledtabular}
\end{table}

\begin{figure}[ht]
   \includegraphics[width=9cm]{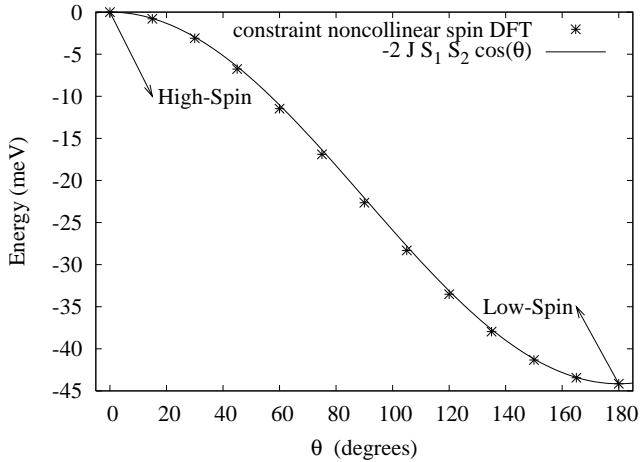}
\caption{LSDA energy change as a function of the angle between the local magnetic moments obtained from a constraint 
 noncollinear spin density functional calculation in the oxovanadium complex (Fig.~\ref{Fig:V2}). 
The solid line shows the (ideal) cosine function 
connecting the AF and FM extrema.}
  \label{Fig:full-LSDA}
\end{figure}

\begin{figure}[ht]
   \includegraphics[width=9cm]{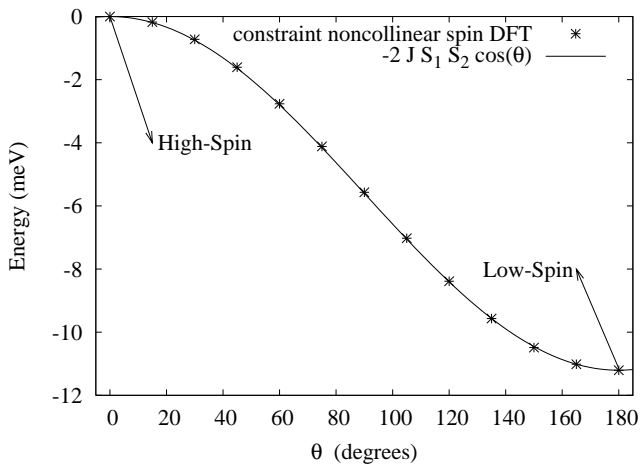}
\caption{Same as Fig.~\ref{Fig:full-LSDA} for B3LYP.}
  \label{Fig:full-B3LYP}
\end{figure}

Our second proof of concept was carried out in the oxovanadium(IV) dimer [($\mu$-OCH$_3$)VO(ma)]$_2$. This complex shows a strong antiferromagnetic 
coupling of about $-$13.3~meV, as measured by temperature-dependent magnetic susceptibility experiments.\cite{V2-exp}
Here we employed Ahlrich's triple-zeta valence basis set for for the V atoms and Ahlrich's double-zeta valence basis for first-row atoms\cite{Ahlrich1,Ahlrich2}, 
as obtained from Ref.~\onlinecite{EMSL}. 
This basis was shown to provide reliable results  in practical calculations of exchange couplings.\cite{Ruiz_basis,Rudra_JCP_124_024103} 
Atomic coordinates were taken from experimental crystallographic data.\cite{V2-exp}
In Figs.~\ref{Fig:full-LSDA}  and \ref{Fig:full-B3LYP} we present a plot of $E^{\sml KS}$ as a function of $\theta$ ($0\le\theta\le180^{\circ}$) 
for LSDA and B3LYP, respectively.
In both figures, $E^{\sml KS}(\theta)$  follows closely a cosine function connecting the HS and LS extrema, indicating that both functionals capture the Heisenberg behavior of 
the oxovanadium complex. Related investigations in periodic systems using the LSDA can be found in the literature.\cite{Kurz_2001,Novak_2008}

\begin{figure}[h]
   \includegraphics[width=7cm]{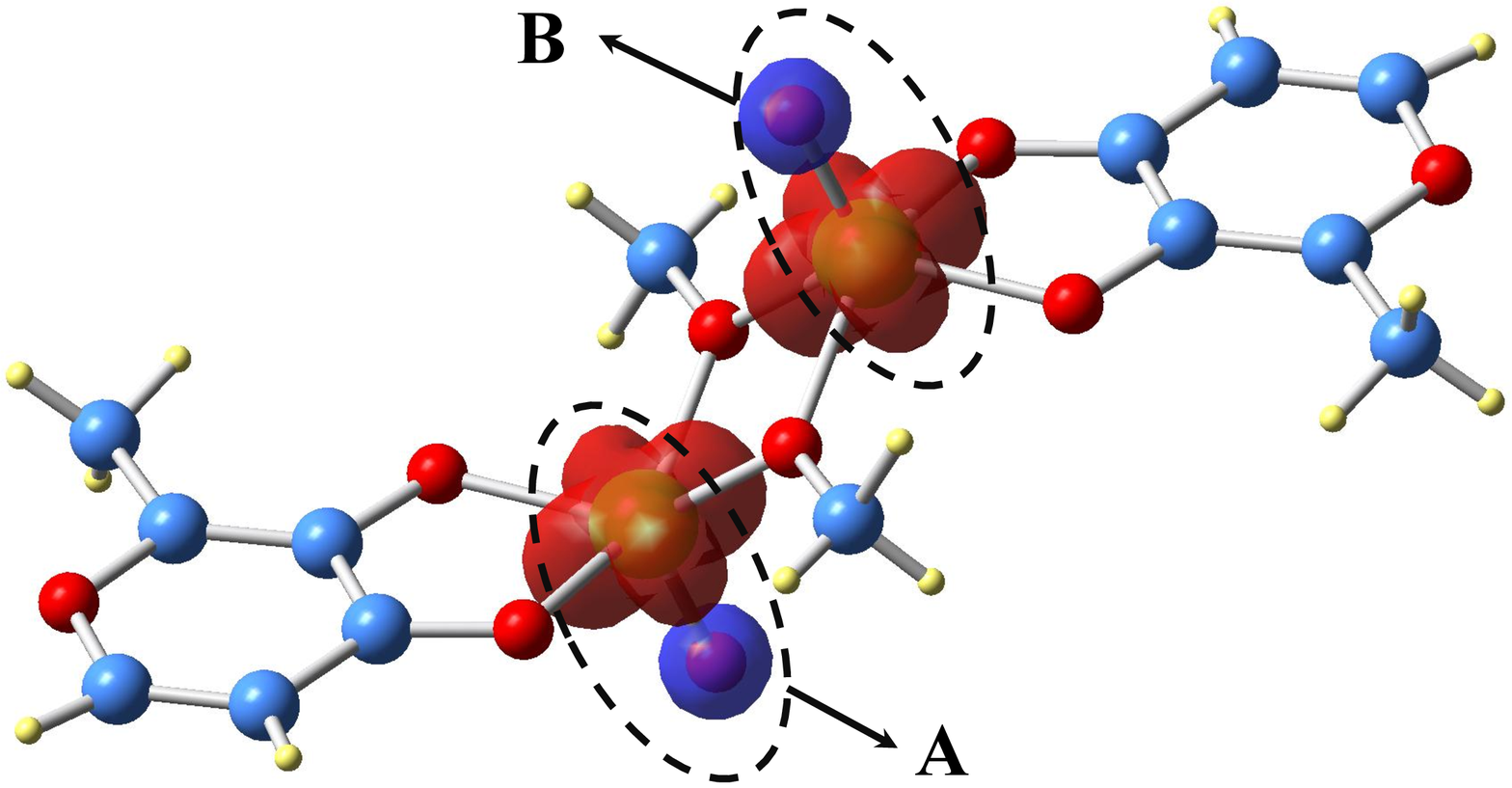} \\
{\bf  (a)}\\
   \includegraphics[width=7cm]{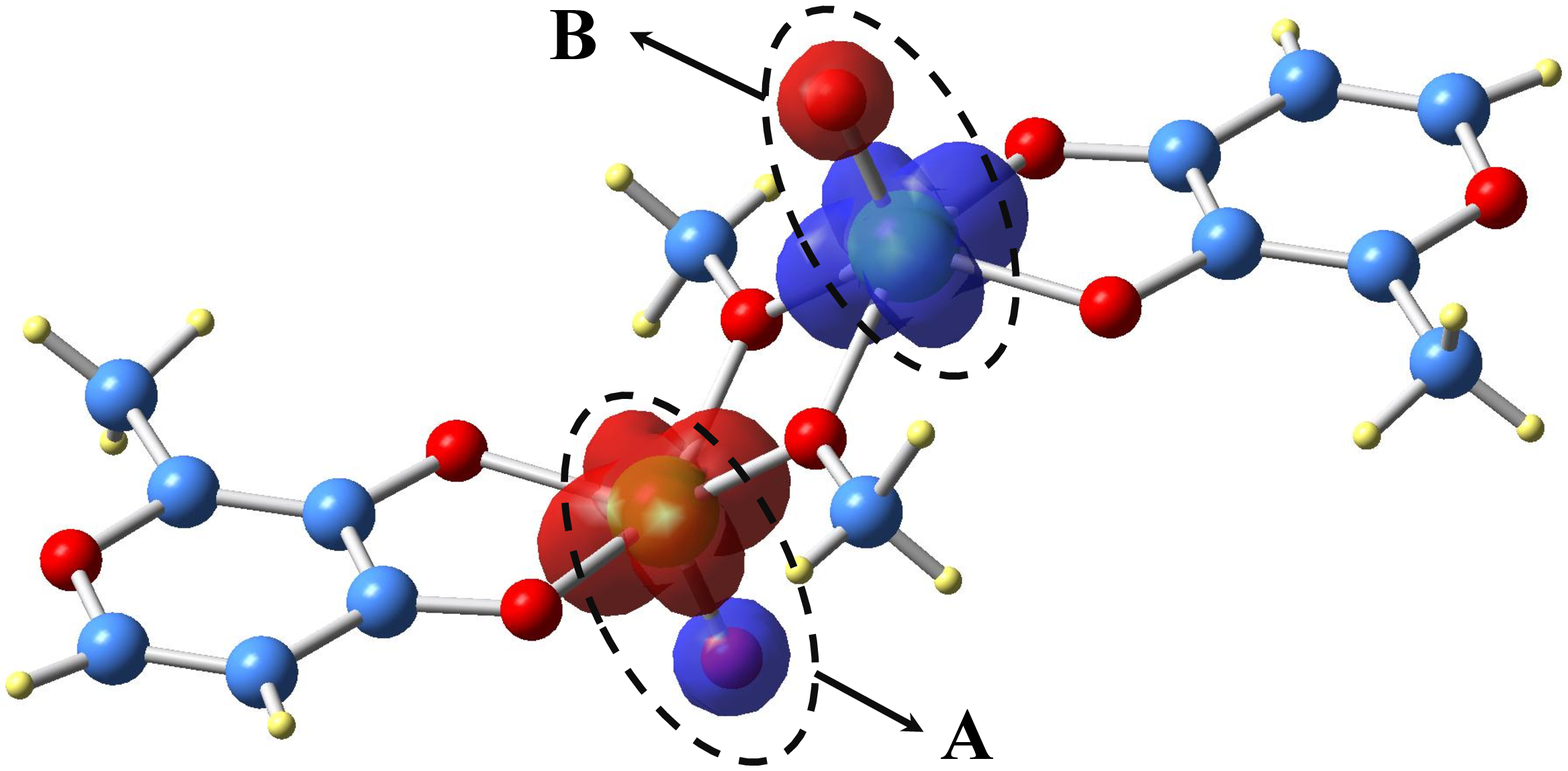} \\
{\bf  (b)}\\
\caption{  Spin density isosurface of the HS (a) and LS (b) states of the oxovanadium complex (Fig.~\ref{Fig:V2}). Red corresponds to $\uparrow$ and blue to $\downarrow$. 
The isosurface represents a spin density of 0.01~(Bohr$^{-3}$). Magnetic centers $A$ and $B$ chosen for the calculations are also indicated. }
\label{Fig:V2}
\end{figure}

For the plots shown in Fig.~\ref{Fig:full-LSDA} and Fig.~\ref{Fig:full-B3LYP} 
we have chosen as magnetic centers $A$ and $B$ ($S_A=S_B=1/2$) both sets of V and apical O atoms since most of the spin density in this complex 
is localized on this moiety, as shown in Fig.~\ref{Fig:V2}. 
However, it is worth to mention that by choosing the metal atoms only as magnetic centers $A$ and $B$  the changes in the plots  are 
unappreciable and the calculated magnetic exchange couplings $J^{\sml HS}$ and $J^{\sml LS}$ vary very little. For instance, for LSDA, 
magnetic exchange couplings change (in meV) from $J^{\sml HS}=-46.6$ to $J^{\sml HS}=-46.9$ 
and from $J^{\sml LS}=-41.9$ to $J^{\sml LS}=-42.2$ when using the V and O atoms or the V atoms only, respectively. 
This indicates that our method is not sensitive upon a particular
choice of the magnetic centers and shows, in this sense, robustness. 
One physical explanation for this fact is that the spin polarization of the light atoms surrounding a metal atom
is ``dragged'' by the  strong magnetic coupling with the neighbor metal
center and 
 therefore, it tends to align parallel (or antiparallel) to the magnetization of the metal center. For the case in study, 
if the constraint is applied on the V atoms only, 
the angle of the spin polarization on the apical O atom deviates from the direction of the constrain vectors by a maximum of 
only 2.75$^{\circ}$ (for $\theta=90^{\circ}$).

\begin{table}
    \caption{
    Magnetic exchange couplings (in meV) calculated from energy derivatives and from energy differences for the vanadium bimetallic complex. 
The corresponding experimental value is  $-$13.3~meV.$^a$
    \label{Table:J-V2}
    }
\begin{ruledtabular}
  \begin{tabular}{lrr}
  & LSDA  &  B3LYP  \\
\cline{1-3}
$J^{\sml HS} = \frac{1}{2 S_A S_B}   \left( \frac{d^2 E}{d\theta^2} \right)_{\theta=0}           $  & $-$46.6   &  $-$11.4  \\[6pt]
$J^{\sml LS} = - \frac{1}{2 S_A S_B} \left( \frac{d^2 E}{d\theta^2} \right)_{\theta=180^\circ} $    & $-$41.9     &  $-$11.2   \\[6pt]
$J^{\sml \Delta E} =  \frac{E(\theta = 180^\circ) - E(\theta = 0)}{4 S_A S_B}$                           & $-$44.1     &  $-$11.3  \\[6pt]
    \end{tabular}
\footnotetext[1]{Taken from Ref.~\onlinecite{V2-exp}}
\end{ruledtabular}
\end{table}

A careful comparison of Fig.~\ref{Fig:full-LSDA} and Fig.~\ref{Fig:full-B3LYP} evidences a larger deviation from the ideal cosine function for LSDA than for B3LYP. 
The largest differences from the cosine function are approximately 0.6~meV and 0.04~meV for LSDA  and B3LYP, respectively, and occurs for $\theta=90^{\circ}$ in both cases. 
This is not surprising since LSDA
yields electron (total and spin) densities more delocalized than B3LYP (the L\"owdin atomic magnetic moments at the V atoms for the LS state 
are 1.00~$\mu_B$ and 1.10~$\mu_B$ for LSDA and B3LYP, respectively) and therefore, 
one can expect that the B3LYP energy follows the Heisenberg behavior 
more closely than its LSDA counterpart.
Localization of the spin-density also reduces the calculated magnetic exchange couplings, as shown by Martin\cite{Martin_and_Illas}, Ruiz\cite{Ruiz_basis,Ruiz_CCR_2005},  
and demonstrated by Rudra et at. by explicitly constraining the 
local magnetization of the LS state.\cite{Rudra_JCP_124_024103} 
As shown in Table~\ref{Table:J-V2}, The difference 
between  $J^{\sml HS} $ and $J^{\sml LS}$ is 4.7~meV for LSDA, while it is only 0.2~meV for B3LYP. Thus, in contrast to 
the perfectly localized case, a more delocalized magnetization yields to 
larger deviations from the ideal Heisenberg behavior and hence greater differences between  $J^{\sml HS} $ and $J^{\sml LS}$ 
and at the same time  to larger
exchange couplings.

\section{Conclusions}

We have proposed a method for the calculation of magnetic exchange couplings from noncollinear spin density functional 
calculations that  employs the second derivative of the electronic energy of a 
single state with respect to a parameter, Eqs.~(\ref{eq:Heis-deriv1}) and (\ref{eq:Heis-deriv2}). Within this approach there is no need to
search for different self-consistent solutions of spin-states as it is commonly done in methods based on energy differences, such as the SP or NP methods, 
Eqs.~(\ref{eq:SP}) and (\ref{eq:NP}). Our method 
utilizes perturbation theory for the evaluation of magnetic exchange couplings and therefore, in 
combination with standard analytic second derivatives techniques, it can potentially be used to compute exchange  couplings very efficiently,
opening the possibility of ``black-boxifying'' the extraction
of magnetic exchange couplings from density functional theory calculations.

Our proof of concept calculations show very promising results. For the cases studied we found  that our method reproduces exchange couplings obtained 
from the spin-projected approach based on  energy differences. As expected from physical grounds, the agreement between 
both methods improves  when the DFT description of the interaction between the 
magnetic centers is more Heisenberg-like. In this sense, 
the curve $E^{\sml KS}(\theta)$ provides a quantitative measure of
how well the  electronic system mimics the behavior of an ideal Heisenberg model.

\section{Acknowledgments}

This research was supported in part by an award from Research Corporation. 
J.E.P. acknowledges support from the President's Research Investment Fund (PRIF)
and a start-up grant from Central Michigan University. 


\end{document}